\documentclass[12pt]{article}
\usepackage{graphicx}
\usepackage[bookmarksnumbered,colorlinks,plainpages]{hyperref}

\begin{document}

\title{Non quantum uncertainty relations of stochastic dynamics}

\author{Qiuping A. Wang\\
{\it Institut Sup\'erieur des Mat\'eriaux et M\'ecaniques Avanc\'es}, \\
{\it 44, Avenue F.A. Bartholdi, 72000 Le Mans, France}}

\date{}

\maketitle

\begin{abstract}
First we describe briefly an information-action method for the study of stochastic
dynamics of hamiltonian systems perturbed by thermal noise and chaotic instability.
It is shown that, for the ensemble of possible paths between two configuration
points, the action principle acquires a statistical form $\langle\delta A\rangle=0$.
The main objective of this paper is to prove that, via this information-action
description, some quantum like uncertainty relations such as $\langle\Delta
A\rangle\geq\frac{1}{\sqrt{2}\eta}$ for action, $\langle\Delta x\rangle\langle\Delta
P\rangle\geq\frac{1}{\eta}$ for position and momentum, and $\langle\Delta
H\rangle\langle\Delta t\rangle\geq\frac{1}{\sqrt{2}\eta}$ for hamiltonian and time,
can arise for stochastic dynamics of classical hamiltonian systems. A corresponding
commutation relation can also be found. These relations describe, through action or
its conjugate variables, the fluctuation of stochastic dynamics due to random
perturbation characterized by the parameter $\eta$.
\end{abstract}


\section{Introduction}
The discussion of this work is limited to mechanical systems without considering the
quantum effect. We use the term {\it regular dynamics} to mean the mechanical
processes with {\it time reversible} trajectories (geodesics) uniquely determined
for each system by the Hamiltonian equations, or equivalently, by least action
principle. Compared to this deterministic character of regular dynamics, one of the
strong difference of irregular (random, stochastic, or statistical) dynamics is the
uncertainty (unpredictability) of the trajectories of the system. This uncertainty
can be illustrated in Figure 1 showing the diffusion of perfume molecules in the
air. Clearly, from the point of view of classical mechanics, if there is no random
perturbation from the molecules of air, an isolated perfume molecule leaving the
hole $a$ of the bottle at time $t_a$ has only one possible path or a fine bundle of
paths having least action and arrives at a sole point $b$ at time $t_b$. With the
random perturbation of thermal noise of air molecules, however, the perfume
molecules can arrive at many very different points $b$ at a time $t_b$. This is the
first dynamical uncertainty due to the random noise. In the theories of chaos, this
uncertainty is usually measured by Kolmogorov-Sinai entropy or by Lyapunov
exponent\cite{Dorfman}. Another uncertainty is that, at time $t_b$, the perfume
molecules can arrive at a point $b$ through different paths. This uncertainty in
path space has been considered, e.g., in the path integral approach to
quantum\cite{Feynman} and non quantum\cite{Wiegel} dynamics and in the large
deviation theory\cite{Oono}. In general, these two uncertainty are not independent
from each other. It is obvious that, without the first uncertainty, the second one
cannot take place.

While the mathematics of regular dynamics can be perfectly formulated on the basis
of the least action principle in classical mechanics, irregular or random dynamics,
found in diffusion, chaotic and other nonequilibrium phenomena, is much more
complicated to be described due to its stochastic and random feature. Nowadays,
statistical and thermodynamic theories of irregular dynamics are still in
development, among which we can cited nonequilibrium statistical
mechanics\cite{Onsager,Kubo}, chaotic dynamics theory\cite{Dorfman}, anomalous
transport theory\cite{Zaslavsky}, large deviation theory\cite{Oono}, small random
perturbation theory\cite{Freidlin} and path integral method starting from Brownian
motion\cite{Wiegel}. Many questions and topics concerning, e.g., time
irreversibility\cite{Mackey}, variational
approaches\cite{Wiegel,Leoncini,Evans,Eyink,Keizer}, chaotic nature of dynamic
process\cite{Cecconi,Mackey2}, connection with quantum
physics\cite{Nelson,Kaniadakis,Nagasawa} are still open to investigation. Much
efforts have been made to clarify the origin of diffusion laws\cite{Bonetto} such as
the Fokker-Planck equation, the Fick's laws, the Fourier law, the Ohm's law and the
anomalous diffusion laws (fractional or nonlinear)\cite{Zaslavsky}.

As a starting point of what we will describe in this paper, we would like to mention
some characteristics shared by large deviation theory, perturbation theory, and path
integral approach. In these theories, variational method is used to find the most
probable (or optimal) paths (histories, trajectories) with the help of rate
functional\cite{Oono}, action functional\cite{Onsager,Freidlin,Eyink,Keizer}, or
path integral\cite{Wiegel,Nagasawa}. All these functionals can be called {\it
effective action} $S$ whose optimization, through a postulated exponential
transition probability with a factor $\exp(-\alpha S)$, allows one to find the most
probable paths. Note that $S$ are not necessarily the mechanistic action defined in
classical mechanics with the Lagrangian $L=E-U$\cite{Wiegel}, where $E$ and $U$ are
respectively kinetic and potential energy. Here $L$ is defined for hamiltonian
systems satisfying (in one dimensional space)
\begin{eqnarray}                                            \label{1}
\dot{x}=\frac{\partial H}{\partial P} \;\;and \;\; \dot{P}=-\frac{\partial
H}{\partial x}
\end{eqnarray}
where $x$ is the coordinates, $P=m\dot{x}$ the momenta, and $H=E+U$ is the
hamiltonian of the system. The mechanistic action is defined by $A=\int_a^b
L(x,P,t)dt$. Its stationary $\delta A=0$ according to least action principle leads
to the Euler-Lagrange equations\cite{Arnold}
\begin{eqnarray}                                            \label{c7x}
\frac{\partial}{\partial t}\frac{\partial L}{\partial \dot{x}}-\frac{\partial
L}{\partial x}=0.
\end{eqnarray}
If we consider the Legendre transformation $H=P\dot{x}-L$, Eqs.(\ref{1}) can be easily
derived from Eq.(\ref{c7x}). It is worth noticing that, when the effective actions
mentioned above are calculated by time integral of an effective Lagrangian,
Eq.(\ref{c7x}) is always satisfied by the optimal
paths\cite{Wiegel,Onsager,Freidlin,Keizer,Nagasawa}.

A point to be noticed is that, in the above theories and their applications, one is
most interested by the most probable paths whose neighborhood provides the basic
contribution to the transition probability\cite{Wiegel,Freidlin,Eyink,Nagasawa}. The
other less probable paths of larger actions are often neglected or only taken into
account (or buried) in the path integrals\cite{Feynman,Wiegel}. As a matter of fact,
the application of classical action principle only to the optimal paths is, to our
opinion, incomplete. The physics represented by the larger action paths is an
inseparable part of the dynamics and may be essential for the fundamental
understanding of irregular dynamics. Recently, an informational method was
proposed\cite{Wang04x,Wang04xx} to treat all the possible paths as an ensemble. The
method considers {\it hamiltonian systems} under random perturbation of thermal
noise and chaotic instability leading to the uncertainties shown in Fig. 1. These
uncertainties can be measured by a path information associated with different paths
between two points in configuration space. The path information is optimized in
connection with the average action of the hamiltonian system calculated over all the
possible paths. We obtain an exponential probability distribution of action. In
principle, this action can be any effective action mentioned above. But in our
previous work, the classical mechanistic action is used. It is worth noticing that
this variational method is proven to be equivalent to an ``extended least action
principle'' for stochastic dynamics, i.e., instead of $\delta A=0$, we have
$\langle\delta A\rangle=0$ where the average $\langle\cdot\rangle$ is taken over all
the paths. Using classical action turns out to be an useful choice since we have
been able to derive the diffusion laws mentioned above for all the possible paths
not only for the optimal paths. In this paper we will describe some mathematical
consequences of this approach. Our main objective here is to derive non quantum
uncertainty relations, which are in fact a necessary consequence of the description
of stochastic dynamics with path probability given by distribution of action.

\section{Optimizing information-action }
First let us recapitulate briefly the method. Let $p_k(b|a)$ be the transition
probability along a path $k$ ($k=1,2,...w$) from a position $a$ and to another
position $b$. The dynamic uncertainty associated with $p_k(b|a)$ is measured with
the following Shannon information
\begin{eqnarray}                                            \label{c1x}
I_{ab}=-\sum_{k=1}^wp_k(b|a)\ln p_k(b|a).
\end{eqnarray}
We have the following normalization $\sum_{k=1}^{w}p_k(b|a)=1$. If $w$ is very large
in an ergodic phase space and if the paths are sufficiently smooth, $\sum_k$ should
be replaced by a path integral in the Feynman sense\cite{Feynman}, e.g.,
$I_{ab}=-\int \mathcal{D}(x)p_k(b|a)\ln p_k(b|a)$ in keeping $a$ and $b$ fixed. The
average action between $a$ and $b$ is given by
\begin{eqnarray}                                            \label{c1xxx}
\langle A_{ab}\rangle=\sum_{k=1}^wp_k(b|a)A_{ab}(k).
\end{eqnarray}
where $A_{ab}(k)$ is the classical action along a path $k$. Our optimal
information-action method consists in the following operation:
\begin{eqnarray}                                            \label{xc1x}
\delta [I_{ab}+\alpha\sum_{k=1}^{w}p_k(b|a)-\eta\sum_{k=1}^wp_k(b|a)A_{ab}(k)]=0
\end{eqnarray}
leading to
\begin{eqnarray}                                            \label{c6x}
p_k(b|a)=\frac{1}{Z}\exp[-\eta A_{ab}(k)],
\end{eqnarray}
where the partition function $Z=\sum_{k}\exp[-\eta A_{ab}(k)]=\int
\mathcal{D}(x)\exp[-\eta A_{ab}(k)]$. It is straightforward to see the following
relationships :
\begin{eqnarray}                                            \label{wc7}
I_{ab}=\ln Z+\eta \langle A_{ab}\rangle
\end{eqnarray}
and
\begin{eqnarray}                                            \label{w7x}
\langle A_{ab}\rangle=-\frac{\partial}{\partial\eta}\ln Z,
\end{eqnarray}
It is proved that\cite{Wang04x} the distribution Eq.(\ref{c6x}) is stable with
respect to the fluctuation of action. If one uses the action of a free
particle\cite{Feynman,Wiegel,Kubo}, $p_k(b|a)$ is just the transition probability of
Brownian particles. In this case, we have a precise physical meaning of the
multiplier $\eta$, i.e., $\eta=\frac{1}{2mD}$ where $m$ is the mass and $D$ the
diffusion constant of the Brownian particle. The signification of $\eta$ will be
discussed later in a general way for particles moving in a potential field $U(x)$.

\section{Extended action principle for irregular dynamics}
As expected, Eq.(\ref{c6x}) is a least action distribution, i.e., the most probable
paths are just the paths of least action $\delta A_{ab}(k)=0$ satisfying
Euler-Lagrange equation and Hamiltonian equations. The other paths do not satisfy
Eqs.(\ref{1}) and (\ref{c7x}). In general, the paths have neither $\delta A_{ab}(k)=0$
nor $\delta \langle A_{ab}\rangle=0$. Eq.(\ref{xc1x}) implies following relationship
\begin{eqnarray}                                            \label{w6x}
-\eta\delta A_{ab}+\delta I_{ab}=0.
\end{eqnarray}
By using Eqs.(\ref{c1x}), (\ref{c1xxx}) and the distribution (\ref{c6x}), it is easy
to calculate that Eq.(\ref{w6x}) is equivalent to
\begin{eqnarray}                                            \label{cx1x}
\langle\delta A_{ab}(k)\rangle=\sum_{k}p_k(b|a)\delta A_{ab}(k)=0.
\end{eqnarray}

On the other hand, in mimicking equilibrium thermodynamics, we can define a dynamic
potential
$$\Psi=\frac{1}{\eta}\ln Z$$
as an analog of the free energy of Helmholtz. From Eqs.(\ref{wc7}) and (\ref{w6x}),
it is straightforward to see that the extended action principle Eq.(\ref{cx1x}) is
equivalent to the following variational principle
\begin{eqnarray}                                            \label{cx1c}
\delta \Psi=0.
\end{eqnarray}
A remarkable application of this above variational approach to computation of
thermodynamic properties of hamiltonian systems was (independently) carried out
recently\cite{Leoncini}. The authors derived thermodynamic equations of motions for
equilibrium systems with different Hamiltonians, already known in the literature and
used in simulations of molecular dynamics, by considering the particle histories in
phase space on constant energy surface.

The extension of action principle has some consequences on the equations of motion as
discussed in \cite{Wang04bx}. For example, for the paths whose action is not at
stationary, we get $\frac{\partial}{\partial t}\frac{\partial L_{k}(t)}{\partial
\dot{x}}-\frac{\partial L_{k}(t)}{\partial x}\neq 0$ and $\dot{P}\neq -\frac{\partial
H}{\partial x}$ which implies a stochastic equation like
\begin{eqnarray}                                            \label{v6x}
\dot{P}_k=-\frac{\partial H}{\partial x}+R
\end{eqnarray}
where $R$ is a random force representing the random perturbation of thermal noise
and chaotic instability. On the other hand, using the action principle
Eq.(\ref{cx1x}), we recover Eqs.(\ref{1}) and (\ref{c7x}) for the ensemble of paths:
\begin{eqnarray}                                            \label{8ab}
\left\langle\frac{\partial}{\partial t}\frac{\partial L_{k}(t)}{\partial
\dot{x}}\right\rangle-\left\langle\frac{\partial L_{k}(t)}{\partial x}\right\rangle=0
\end{eqnarray}
and
\begin{eqnarray}                                            \label{1ab}
\left\langle\dot{x}\right\rangle=\left\langle\frac{\partial H}{\partial
P}\right\rangle \;\;and \;\;
\left\langle\dot{P}\right\rangle=-\left\langle\frac{\partial H}{\partial
x}\right\rangle.
\end{eqnarray}
This implies that the mean of the ``random force'' $R$ over all possible paths must
vanish, i.e., $\left\langle R\right\rangle=0$, required by the extended action
principle. The second equality of Eqs.(\ref{1ab}) is the random analog of Newton's
law and has the same content as the Feynman-Hibbs quantum Newton's law in Eq.(7-42)
of \cite{Feynman}.

In what follows, we are concerned with the spreads of the distribution of the
stochastic dynamics, in other words, the deviation of the irregular dynamics from the
regular one due to random perturbation. This uncertainty is a priori measured by the
path information we introduced. But here it will be analyzed at the level of
mechanical quantity like position, momentum, action and energy.

\section{Uncertainty relations}
For the sake of simplicity, we suppose the point $b$ at $x$ is very close to the
point $a$ at $x_0$ and the transition takes place in the infinitesimal time interval
$\delta t=t-t_0$. This segment can be considered as a factor (a small element of a
long path) in path integral technique\cite{Feynman}. Let $\delta x=x(t)-x_0(t_0)$,
the action is given by
\begin{eqnarray}                                            \label{xx9c}
A_{ab}(x)=\frac{m(\delta x)^2}{2\delta t} +F\frac{\delta x}{2}\delta t-U(x_0)\delta
t,
\end{eqnarray}
where $F=-\left(\frac{\partial U}{\partial x}\right)_{(x+x_0)/2}$ is the force on
the path and $m$ the mass of the studied system. The transition probability is
\begin{eqnarray}                                            \label{xxxc9}
p_k(b|a) &=& \frac{1}{Z} \exp\left(-\eta\left[\frac{m}{2\delta t}\delta x^2
+F\frac{\delta t}{2}\delta x\right]\right)
\end{eqnarray}
with
\begin{eqnarray}                                            \label{xxx9}
Z&=&\int_{-\infty}^\infty dx\exp\left(-\eta\left[\frac{m}{2\delta t}\delta x^2
+F\frac{\delta t}{2}\delta x\right]\right)\\ \nonumber &=& \exp\left[F^2\frac{\eta
\delta t^3}{8m}\right]\sqrt{\frac{2\pi \delta t}{m\eta}}.
\end{eqnarray}
The potential energy of the point $x_0$ disappears after normalization because it does
not depend on $x$. $F$ is considered constant on small $\delta x$. It is easy to
show\cite{Wang04xx} that the probability of Eq.(\ref{xxxc9}) satisfies the
Fokker-Planck equation and that other diffusion laws can be trivially derived from it.

How far are the randomly perturbed paths deviated from the optimal paths of regular
dynamics? How different are Eqs.(\ref{1ab}) from Eqs.(\ref{1})? This question can be
answered, under different angles, by the standard deviation of action $\langle\Delta
A\rangle^2=\langle A^2\rangle-\langle A\rangle^2=\langle A^2\rangle-\langle
A_{ab}\rangle^2=-\frac{\partial \langle A_{ab}\rangle}{\partial \eta}$, of position
$\langle\Delta x\rangle^2=\langle x^2\rangle-\langle x\rangle^2$, of momentum
$\langle\Delta P\rangle^2=\langle P^2\rangle-\langle P\rangle^2$ and of Hamiltonian
$\langle\Delta H\rangle^2=\langle H^2\rangle-\langle H\rangle^2$. Using the formula
below\footnote{$\int_{-\infty}^\infty dx\exp(-\alpha x^2+2\gamma
x)=\exp(\gamma^2/\alpha)\sqrt{\frac{\pi}{\alpha}}$ and $\int_{-\infty}^\infty
x^ndx\exp(-\alpha x^2+2\gamma
x)=\frac{1}{2^{n-1}\alpha}\sqrt{\frac{\pi}{\alpha}}\frac{d^{n-1}}{d\gamma^{n-1}}
[\gamma\exp(\gamma^2/\alpha)]$.}, we obtain
\begin{eqnarray}                                            \label{6a0}
\langle x-x_0\rangle=-\frac{F\delta t^2}{2m},
\end{eqnarray}
\begin{eqnarray}                                            \label{6a1}
\langle P-P_0\rangle=-\frac{F\delta t}{2},
\end{eqnarray}
\begin{eqnarray}                                            \label{6a2}
\langle (x-x_0)^2\rangle=\frac{\delta t}{m\eta}+\frac{F^2\delta t^4}{4m^2},
\end{eqnarray}
and
\begin{eqnarray}                                            \label{6a3}
\langle (P-P_0)^2\rangle=\frac{m}{\delta t\eta}+\frac{F^2\delta t^2}{4}.
\end{eqnarray}
Since $\langle\Delta x\rangle=\langle\Delta (x-x_0)\rangle$, the above relationships
imply
\begin{eqnarray}                                            \label{6a4}
\langle\Delta x\rangle^2=\frac{\delta t}{m\eta}
\end{eqnarray}
and
\begin{eqnarray}                                            \label{6a5}
\langle\Delta P\rangle^2=\frac{m}{\delta t\eta}.
\end{eqnarray}
Cancelling the time $\delta t$ and $m$ in the above two equations, we get a
``classical uncertainty relation'' for irregular dynamics
\begin{eqnarray}                                            \label{6a6}
\langle\Delta x\rangle^2\langle\Delta P\rangle^2=\frac{1}{\eta^2}.
\end{eqnarray}
Notice that this is only an asymptotic relation for $\delta t\rightarrow 0$. For a
measurable (longer) length of time and path, we divide them into small segments of
the order of $\delta t$ and $\delta x$. Eq.(\ref{6a6}) should be valid for each
segment. Thus it can be trivially proven that the total deviation
$\langle\Delta\rangle^2$ of whatever quantity on a path is a sum of all the
deviations on its small segments (Gaussian law of errors, see for example
\cite{Kubo}). So in general, we must write
\begin{eqnarray}                                            \label{a6}
\langle\Delta x\rangle^2\langle\Delta P\rangle^2\geq\frac{1}{\eta^2}
\end{eqnarray}
which is valid for any measurable period of time and path length. If the system is
in rotation with $\delta x/R=\delta \theta$ where $\theta$ is the rotation angle and
$1/R$ the curvature of $\delta x$, then we get
\begin{eqnarray}                                            \label{aa6}
\langle\Delta \theta\rangle^2\langle\Delta J\rangle^2\geq\frac{1}{\eta^2},
\end{eqnarray}
where $J$ is the angular momentum.

The action uncertainty can be calculated from Eqs.(\ref{w7x}) and (\ref{xxx9}) :
\begin{eqnarray}                                            \label{6a7}
\langle A_{ab}\rangle=\frac{1}{2\eta}-\frac{F^2\delta t^3}{8m}
\end{eqnarray}
so that
\begin{eqnarray}              \nonumber                    \label{6a8}
\langle\Delta A\rangle^2 &=& -\frac{\partial \langle A_{ab}\rangle}{\partial \eta} \\
&=& \frac{1}{2\eta^2}.
\end{eqnarray}
This relation can be written as $\langle\Delta L\rangle^2\langle\delta
t\rangle^2=\frac{1}{2\eta^2}$ where $\langle\Delta L\rangle^2=\langle
L^2\rangle-\langle L\rangle^2$ is the standard deviation of Lagrangian. For
arbitrary time and path length, we must write
\begin{eqnarray}                                 \label{6aa8}
\langle\Delta A\rangle^2 \geq\frac{1}{2\eta^2}.
\end{eqnarray}
and
\begin{eqnarray}                                        \label{6a8a}
\langle\Delta L\rangle^2\langle\Delta t\rangle^2\geq \frac{1}{2\eta^2}
\end{eqnarray}
where $\langle\Delta t\rangle^2=\langle t^2\rangle-\langle t\rangle^2$ is the standard
deviation of time measure. Eq.(\ref{6a8a}) can be verified in the same way as for
position and momentum, through the calculation of $\langle t\rangle$ and $\langle
t^2\rangle$ with the distribution Eq.(\ref{xxxc9}) in relaxing $\delta t$ and fixing
$\delta x$.

Now let $H(t)$ be the Hamiltonian on the considered path segment at moment $t$, we
have the Legendre transformation $H(t)=P\dot{x}-L(t)$. Considering $L=E-U$ and
$E=P\dot{x}/2$, we get
\begin{eqnarray}                                            \label{6a9}
\langle\Delta H\rangle^2=\langle\Delta L\rangle^2+4(\langle EU\rangle-\langle
E\rangle\langle U\rangle).
\end{eqnarray}
$E$ and $U$ are two independent variable, so $\langle EU\rangle=\langle
E\rangle\langle U\rangle$, we obtain
\begin{eqnarray}                                             \label{6a10}
\langle\Delta H\rangle^2=\langle\Delta L\rangle^2
\end{eqnarray}
and
\begin{eqnarray}                                             \label{a8a}
\langle\Delta H\rangle^2\langle\Delta t\rangle^2\geq \frac{1}{2\eta^2}.
\end{eqnarray}

Before giving an interpretation of these non quantum uncertainty relations of
irregular dynamics, i.e., Eq.(\ref{a6}), Eq.(\ref{aa6}), Eq.(\ref{6aa8}),
Eq.(\ref{6a8a}) and Eq.(\ref{a8a}), we would like to indicate two points. First, it
seems that Eq.(\ref{6aa8}), the action uncertainty relation, is the most essential one
because it tell us the spread of the action distribution and the deviation of the
extended action principle $\langle\delta A\rangle=0$ from the conventional one $\delta
A=0$. The other relations concerns the conjugate variables of action and can be a
priori derived from action uncertainty. Second, the coefficient $\eta$ plays a central
role in this description of the dynamics. Roughly speaking, $\eta\rightarrow 0$
represents large deviation of the perturbed dynamics from the regular one (large
uncertainty), large $\eta$ represents small random perturbation and deviation (small
uncertainty), and $\eta\rightarrow\infty$ implies vanishing perturbation and
convergence of irregular dynamics to regular one (zero uncertainty).

A quantitative relationship between the uncertainty and thermal noise can be shown
with a special example, Brownian motion ($F=0$) with diffusion constant $D=\mu
k_BT$\cite{Kubo}, where $\mu$ is the mobility of the particles, $k_B$ is the
Boltzmann constant and $T$ the temperature. Here we see a relationship between
$\eta$ and temperature characterizing the thermal noise: $\eta=\frac{1}{2m\mu
k_BT}$. In this case, we can write, for example,
\begin{eqnarray}                                            \label{a6a6}
\langle\Delta x\rangle\langle\Delta P\rangle\geq 2m\mu k_BT,
\end{eqnarray}
which implies that the paths may be distributed very widely in phase space around
the optimal (least action) paths at high temperature.

\section{A commutation relation}
Now let us define a certain functional $G_k(x)$ of paths. The average of the
functional over all the paths is given by
\begin{eqnarray}                                            \label{cc6x}
\langle G_k(x)\rangle=\sum_kG_k(x)p_k(b|a).
\end{eqnarray}
Now differentiate this equation with respect to $x$, the average does not depend on
position, so $\frac{\partial \langle G_k(x)\rangle}{\partial x}=0$. The derivative
of right hand side yields two terms: $\sum_kp_k(b|a)\frac{\partial G_k(x)}{\partial
x}$ and $-\eta\sum_kp_k(b|a)G_k(x)\frac{\partial A_{ab}(k)}{\partial x}$. This
implies
\begin{eqnarray}                                            \label{cc6xx}
\left\langle \frac{\partial G_k(x)}{\partial x}\right\rangle=-\eta \left\langle
G_k(x)\frac{\partial A_{ab}(k)}{\partial x}\right\rangle,
\end{eqnarray}
which is the classical analog of the Feynman-Hibbs relation given by Eq.(7-30) of
\cite{Feynman}. Imposing $G_k(x)=x$, we obtain\cite{Feynman}:
\begin{eqnarray}                                            \label{ac6xx}
\left\langle m\frac{x_{i+1}-x_i}{\delta t}x_i\right\rangle-\left\langle
x_im\frac{x_{i}-x_{i-1}}{\delta t} \right\rangle=\frac{1}{\eta},
\end{eqnarray}
where $i$ is the index of a segment for the time period $\delta t=t_i-t_{i-1}$ of a
certain path $k$. This relation can be written as
\begin{eqnarray}                                            \label{6xxc}
\left\langle P_{i+1}x_i\right\rangle-\left\langle x_iP_i
\right\rangle=\frac{1}{\eta},
\end{eqnarray}
which can be considered as a classical commutation relation, where $x_i$ is the
position at time $t_i$, $P_{i+1}=\frac{x_{i+1}-x_i}{\delta t}$ and
$P_{i}=\frac{x_{i}-x_{i-1}}{\delta t}$ are the momenta at time $t_{i+1}$ and $t_i$,
respectively.

We would like to indicate that the above results, i.e., the uncertainty and
commutation relations, were obtained by using the mechanical Lagrangian and action.
Nevertheless, as mentioned above, the same conclusions can be reached with other
action functionals and Lagrangians. It can be trivially shown that similar
uncertainty relations arise with, e.g., Onsager-Machlup Lagrangian
$L_{OM}=(\dot{x}-\mu F)^2$\cite{Wiegel,Onsager} and Freidlin-Wentzell Lagrangian
$L_{OM}=[\dot{\varphi}-b(\varphi)]^2$ where $\dot{\varphi}=b(\varphi)$ is the
differential equation of some continuous function $\varphi$ of paths when the
perturbation vanishes\cite{Freidlin}. If the Lagrangians do not have mechanistic
interpretation, the uncertainty relations should be understood differently depending
on the physical content of the variables.

\section{Concluding remarks}
We have briefly presented an information-action method for stochastic dynamics of
hamiltonian systems. We see that in this approach the Euler-Lagrange equations, the
Hamiltonian equations and the action principle only hold in averaged form over the
ensemble of all the possible paths between two state points. The main result of this
paper is to show that some uncertainty relations, very similar to those in quantum
mechanics, exist for stochastic dynamics of hamiltonian systems.

We are in the realm of classical physics, no quantum effect is considered. In
addition, the systems under consideration are possibly macroscopic. There is
normally the consensus that exact value of position and speed can be a priori
assigned to a particle at the same time. Then how to understand the uncertainty
relations in classical dynamics? A plausible understanding is that $\langle\Delta
A\rangle\geq\frac{1}{\sqrt{2}\eta}$ implies the distribution of action cannot be
arbitrarily narrow around the least action, or the fluctuation of action has a non
zero minimum due to the random perturbation. In the same way, $\langle\Delta
x\rangle\langle\Delta P\rangle\geq\frac{1}{\eta}$ means that the fluctuations or the
distributions of position and momentum cannot be $simultaneously$ arbitrarily
narrow. It has no meaning about the precision of the measure, because a classical
body does have exact position and speed and the means $\langle x\rangle$ and
$\langle P\rangle$ can be made arbitrarily close to the real values of the body
whatever the fluctuation $\langle\Delta x\rangle$ and $\langle\Delta P\rangle$ of
the separately measured values.

A question naturally arises. Do these classical uncertainty relations have something
to do with their quantum analogs? Remember that in this work we consider classical
systems under the random perturbation of thermal noise. This essential point is
illustrated by the example of Brownian motion in Eq.(\ref{a6a6}) which tell us that
if the thermal noise vanishes for $T\rightarrow 0$, we get, e.g., $\langle\Delta
x\rangle\langle\Delta P\rangle\geq 0$. So no quantum effect can be observed here.
One needs $additional$ hypothesis in order to enter into the quantal world from the
present results. For example, if one assumes $\eta=1/\hbar$, the distribution
function Eq.(\ref{c6x}) becomes an {\it universal Brownian distribution} of
F\'enyes-Nelson\cite{Nelson,Fenyes} satisfying Fokker-Planck equation\cite{Wang04bx}
whose forward and backward versions can lead to Schrödinger
equation\cite{Nelson,Nagasawa}. At the same time the classical uncertainty relations
become the Heisenberg's relations. To our opinion, the passage from classical
stochastic dynamics to the stochastic (quantum) mechanics of F\'enyes-Nelson is not
automatic. The assumptions of an universal noise and the emergence of a constant
minimum uncertainty $\hbar$ are not trivial. A priori, without any hypotheses, the
results of the present work are meaningful only for classical systems.

The idea of classical uncertainty relations can be seen in the interpretation of
quantum mechanics on the basis of the hypothesis of Cantorian (fractal)
space-time\cite{Naschie,Nottale,Ord,Jumarie,Boyarsky}. Some authors discussed the
quantum uncertainty relations due to fractal quantum
paths\cite{Nottale,Ord,Naschie2,Naschie3}. Their reasoning is in fact extendable to
macroscopic non quantum fractal paths as discussed in \cite{Nottale,Ord}. This is
the case of classical uncertainty relations arising from fractal geometry.

As a matter of fact, a possible classical background of the Heisenberg's relations
was proposed long ago by F\"urth\cite{Furth} and then discussed by many
auhtors\cite{Fenyes,Guth}. There the authors investigated an uncertainty relation of
Brownian motion for position and drift velocity of diffusion, the latter is not the
instantaneous velocity (or momentum), i.e., the conjugate variable of position used
in Eq.(\ref{6a6}). It was believed that\cite{Guth} the Bohr's concept of
complementarity also enters in classical mechanics, but with a statistical average
of momentum; this implies that there is no classical analogs of quantum commutation
relations. The present work shows that the classical analogs of quantum relations
exist.

\newpage

\begin{figure}[h] \label{f1}
\includegraphics[width=14cm,height=16cm]{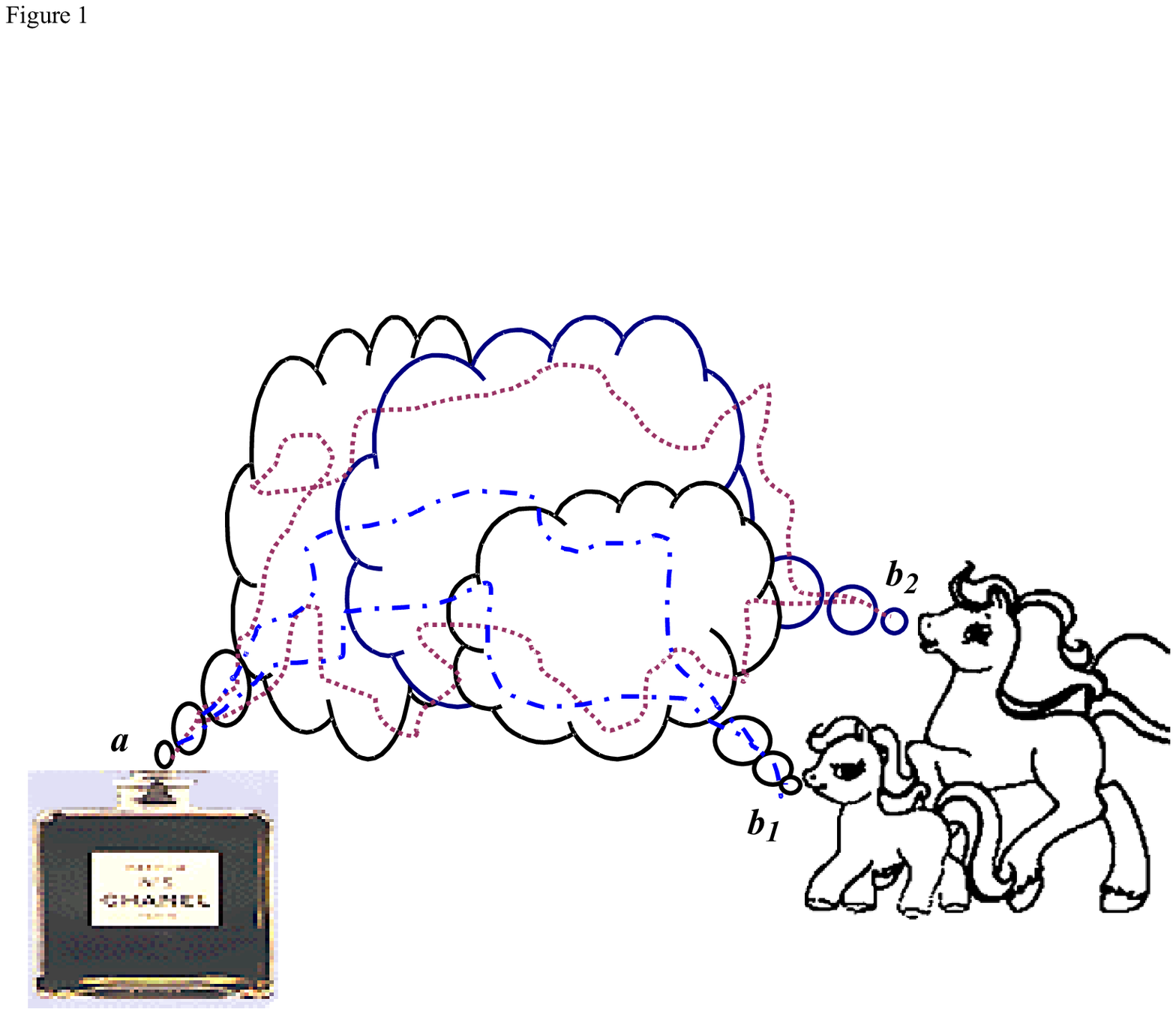}
\caption{An example of random dynamics: the diffusion of scent. At time $t_a$, the
molecules of the perfume get out of the bottle at point $a$. At time $t_b$, the
molecules arrive at different points $b_i$ ($i=1,2,...$) (first dynamical
uncertainty). For a pony putting his nose at a given point $b$, all the molecules it
receives at time $t_b$ may arrive there via different paths (second dynamical
uncertainty). Obviously, as shown by the dotted lines in the figure, the second
uncertainty cannot take place without the first one. }
\end{figure}
\newpage

\end{document}